\documentclass[reprint,amsmath,amssymb,aps,superscriptaddress]{revtex4-1}

\usepackage{graphicx}
\usepackage{dcolumn}
\usepackage{bm}
\usepackage{multirow}

\begin{document}

\title{Long distance co-propagation of quantum key distribution and  terabit  classical optical data channels}

\author{Liu-Jun Wang}
\affiliation{Hefei National Laboratory for Physical Sciences at Microscale and Department of Modern Physics, University of Science and Technology of China, Hefei, Anhui 230026, China\\
}
\affiliation{CAS Center for Excellence and Synergetic Innovation Center in Quantum Information and Quantum Physics, University of Science and Technology of China, Hefei, Anhui 230026, China}

\author{Kai-Heng Zou}
\affiliation{State Key Laboratory of Advanced Optical Communication Systems and Networks, Peking University, Beijing 100871, China}

\author{Wei Sun}
\author{Yingqiu Mao}
\affiliation{Hefei National Laboratory for Physical Sciences at Microscale and Department of Modern Physics, University of Science and Technology of China, Hefei, Anhui 230026, China\\
}
\affiliation{CAS Center for Excellence and Synergetic Innovation Center in Quantum Information and Quantum Physics, University of Science and Technology of China, Hefei, Anhui 230026, China}

\author{Yi-Xiao Zhu}
\affiliation{State Key Laboratory of Advanced Optical Communication Systems and Networks, Peking University, Beijing 100871, China}

\author{Hua-Lei Yin}
\affiliation{Hefei National Laboratory for Physical Sciences at Microscale and Department of Modern Physics, University of Science and Technology of China, Hefei, Anhui 230026, China\\
}
\affiliation{CAS Center for Excellence and Synergetic Innovation Center in Quantum Information and Quantum Physics, University of Science and Technology of China, Hefei, Anhui 230026, China}

\author{Qing Chen}
\author{Yong Zhao}
\affiliation{QuantumCTek Co., Ltd., Hefei, Anhui 230088, China}

\author{Fan Zhang}
\email{fzhang@pku.edu.cn}
\affiliation{State Key Laboratory of Advanced Optical Communication Systems and Networks, Peking University, Beijing 100871, China}

\author{Teng-Yun Chen}
\email{tychen@ustc.edu.cn}
\author{Jian-Wei Pan}
\email{pan@ustc.edu.cn}
\affiliation{Hefei National Laboratory for Physical Sciences at Microscale and Department of Modern Physics, University of Science and Technology of China, Hefei, Anhui 230026, China\\
}
\affiliation{CAS Center for Excellence and Synergetic Innovation Center in Quantum Information and Quantum Physics, University of Science and Technology of China, Hefei, Anhui 230026, China}

\begin{abstract}
Quantum key distribution (QKD) generates symmetric keys between two remote parties, and guarantees the keys not accessible to any third party. Wavelength division multiplexing (WDM) between QKD and classical optical communications by sharing the existing fibre optics infrastructure is highly desired in order to reduce the cost of QKD applications. However, quantum signals are extremely weak and thus easily affected by the spontaneous Raman scattering effect from intensive classical light. Here, by means of wavelength selecting and spectral and temporal filtering, we realize the multiplexing and long distance co-propagation of QKD and Terabit classical coherent optical communication system up to 80km. The data capacity is two orders of magnitude larger than the previous results. Our demonstration verifies the feasibility of QKD and classical communication to share the resources of backbone fibre links, and thus taking the utility of QKD a great step forward.
\end{abstract}

\maketitle

\section{Introduction}
Quantum key distribution (QKD) \cite{Bennett84,EKERT91,Gisin02} supplies information-theoretic security \cite{Scarani09} based on the principles of quantum mechanics. Since its introduction in 1984 \cite{Bennett84}, QKD has undergone dramatic progress from point-to-point experiments \cite{Takesue07,Stucki09,Dixon10,Liu10,Wang12,Tanaka12,korzh2015provably} to QKD network implementations \cite{Peev09,Chen10,Sasaki11,Froehlich13}. Due to its extremely low intensity, conventional QKD signals require an exclusive fibre link dedicated to its transmission. To avoid the high cost of laying extra fibre resources, the integration of QKD with conventional telecom fibre channels is of great importance. One popular solution is to multiplex QKD with classical optical channels through wavelength division multiplexing (WDM), which was first realized by Townsend in 1997 \cite{Townsend97}, and further extended to optical access and metropolitan networks with moderate classical bit rate of Gigabit per second \cite{Chapuran09,Choi11,Choi10,Peters09,lancho2009qkd,mora2012simultaneous,Aleksic13,
Ciurana14,aleksic2015perspectives}. So far, QKD has not been multiplexed into high data capacity optical backbone links. In this paper, we demonstrate for the first time that QKD can be deployed in Terabit classical optical communication environments with long distance fibre link up to 80km, which shows the integration feasibility of QKD and classical telecom backbone infrastructure.

Classical backbone links have characteristics of long distances and high throughputs. For instance, a typical span distance is 80 km, and the communication capacity of one fibre link reaches up to Terabits per second (Tbps) magnitude. Unfortunately, the highest experimental and field trial record of classical data channel bandwidth used to simultaneously transmit QKD is 40 Gigabits per second (Gbps) \cite{Patel14,choi2014field}. In fact, from the simulation results of Patel \textit{et al}. \cite{Patel14}, when quantum signals wavelength is located at C-band (1530 - 1565 nm), the maximum bandwidth of data channels achievable was predicted to be 140 Gbps. This is because that as the bandwidth increases, the classical light launch power also increases, resulting in stronger spontaneous Raman scattering noise and linear crosstalk induced by the classical light, which are the main obstacles in WDM integration of QKD and classical optical data channels.

Besides, in previous experiments the classical communication generally used on-off keying (OOK) modulation schemes, which is intensity modulated and detected directly by photodiodes. With this kind of modulation, a bit rate of 1 Gbps typically corresponds to a launch power of 0 dBm (1 mW). As the bit rate is basically proportional to the launch power, 1 Tbps OOK data communication would require 30-dBm classical light, which will result in the unacceptably severe Raman scattering noise. Fortunately, the Tbps classical data channels are currently implemented by coherent optical communication combined with M-ary quadrature amplitude modulation (QAM) formats. By using 16-QAM and 64-QAM in our experiment, the classical optical power is about 10 dBm at Tbps level, which provides the possibility of QKD multiplexing. We note that high-order QAMs require higher optical signal to noise ratios (OSNR) than OOK modulations. The low launch power will lead to a worse OSNR while the high launch power will result in severe fibre nonlinear distortions that deteriorate the signal quality. Therefore, for a specific transmission distance, there exists an optimum launch power as a trade-off to balance the influence of noise and nonlinear interference.

There are some points of consideration when multiplexing QKD with such a high capacity classical optical communication. Firstly, we need to suppress the in-band noise that has the same wavelength as our quantum signals, which comes from the background fluorescence of the classical light source and the amplified spontaneous emission noise generated from erbium-doped fibre amplifiers (EDFA). Secondly, we require a high degree of isolation to reduce the out-of-band noise, which corresponds to the probability of classical light being detected by the single-photon detectors at the QKD receiver site. These two kinds of noise are proportional to the incident light power and generally referred to as the linear crosstalk. In fact, the main challenge of WDM comes from the spontaneous Raman scattering effect from the intensive classical light \cite{auyeung1978spontaneous,Subacius05,xavier2009scattering,da2014impact}.

Here, we show that through wavelength selection and sharp optical filtering, the multiplexing between QKD and Terabit classical data channels can be successfully achieved. Such coexistence leverages the existing backbone fibre cables, realizing large cost savings potentials over deploying dedicated quantum links.

\section{Raman noise and secure key rates at 1550.12 \lowercase{nm} and 1310 \lowercase{nm}}
In order to quantify the impact of Raman scattering on QKD, we first need to determine the classical signal wavelength $\lambda_c$ and quantum signal wavelength $\lambda_q$. The commercial dense WDM (DWDM) technology usually uses the C-band with relatively low fibre loss. Note that EDFAs are generally necessary to compensate for the fibre link attenuation. In contrast, the quantum signals cannot be amplified in principle because of the no-cloning theorem  \cite{wootters1982single,dieks1982communication}. Therefore, in previous point-to-point WDM experiments, $\lambda_c$ was usually chosen in the C-band, while $\lambda_q$ was also located at C-band because of its low fibre loss \cite{xia2006band,Eraerds10,Patel12,walenta2014fast,wang2015experimental}, or at the O-band (1260 - 1360 nm) because of its low Raman noise \cite{runser2005demonstration,Nweke05,Chapuran09,Choi11}. Hence, we need to consider the two factors together to determine the appropriate quantum signal wavelength in different classical optical communication environments.

\begin{figure}[tbh]
\centering
\includegraphics[width=1\linewidth]{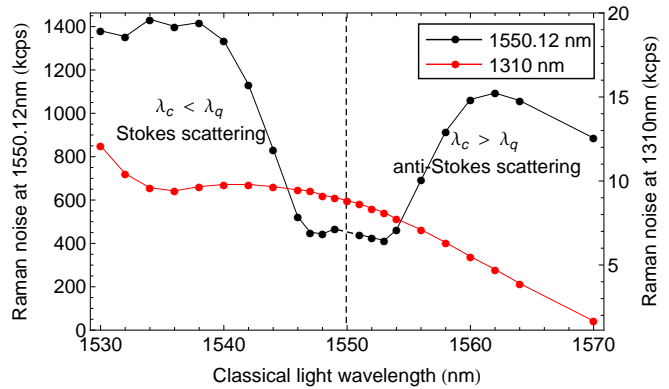}
\caption{Raman noise at 1550.12 nm (black dots) and 1310 nm (red dots). The forward Raman noises are measured in kilo counts per second (kcps) as a function of classical light wavelength, in 13.6 km standard single-mode fibre at room temperature. The classical launch power is 6 dBm.}
\label{fig:spectrum1}
\end{figure}

We measure the forward Raman scattering noises, which transmit in the same direction as the incident light, at both 1550.12 nm and 1310 nm using an InGaAs avalanche photodiode (APD) based single-photon detector, operating at 1.25 GHz with a 180-ps full width at half maximum (FWHM) gate width. Figure~\ref{fig:spectrum1} shows the count rate of Raman noise generated from a continuous wave laser source tuned from 1530 nm to 1570 nm and launched with a power level of 6 dBm. We note that in the $\lambda_q$=1550.12 nm configuration, the QKD receiver used a 20-GHz fibre Bragg gating (FBG) to filter the Raman noise, which induces an extra loss of 3.2 dB. While in the $\lambda_q$=1310 nm configuration, a bandpass filter with center wavelength of 1310.0 nm, passband width of 100 GHz, and insertion loss of 0.5 dB was used. Consequently, the Raman noise at 1550.12 nm strongly depends on the incident light wavelength, which has a count rate of 440.4 kilo counts per second (kcps) on average between 1550.12 $\pm$ 3 nm, and two times more counts beyond 1550.12 $\pm$ 10 nm. Moreover, Fig. 1 shows the intensity of the anti-Stokes scattering slightly weaker than that of the Stokes scattering. Meanwhile, the averaged noise count rate at 1310 nm is 6.2 kcps, and decreases slightly with the increasing classical signal wavelength. We can see that although the received bandwidth of 1550.12 nm is 1/5 of that of 1310 nm, the Raman noise at 1550.12 nm is approximately two orders of magnitude higher than that of 1310 nm. Nevertheless, typical fibre attenuation at 1310 nm is 0.33 dB/km, which is larger than the loss of 0.2 dB/km at 1550.12 nm.

\begin{figure}[tbh]
\centering
\includegraphics[width=\linewidth]{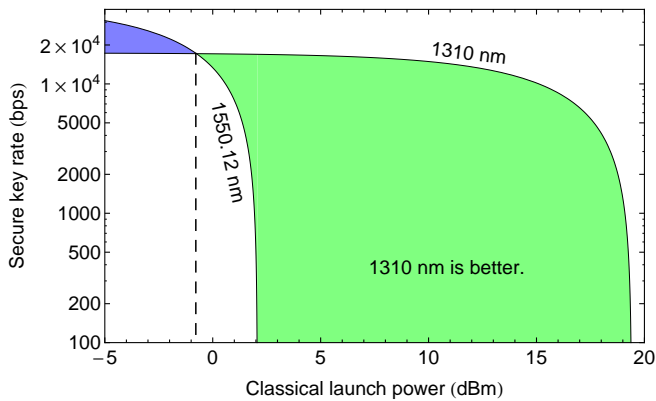}
\caption{Secure key rate comparison between 1550.12 nm and 1310 nm. The secure key rates were calculated at 50 km fibre length as a function of classical launch power. The blue region indicates that 1550.12 nm offers a higher key rate than 1310 nm, while the green area indicates vice versa and that 1310 nm is more suitable for the quantum signal wavelength in a Tbps environment, where the classical launch power is around 10 dBm.}
\label{fig:KeyRateCompare}
\end{figure}

In order to compare the secure key rates of the two quantum signal wavelengths, we consider a scenario of QKD co-propagating with classical channels in a 50 km fibre, and simulate the key rate as a function of classical launch power as shown in Fig.~\ref{fig:KeyRateCompare}. The Raman scattering coefficient we used is obtained from the measured data of Raman noise in Fig.~\ref{fig:spectrum1}, and the QKD key rate simulation follows the decoy method \cite{Ma05}. One can see that the key rate corresponding to $\lambda_q$=1550.12 nm is higher than that of 1310 nm when the classical launch power is less than -0.76 dBm, because the Raman noise for both cases is small at low classical launch power, while 1550.12 nm has the advantage of low fibre loss.

As the optical launch power increases, the disturbance from Raman noise becomes evident that it deteriorates quantum signals of 1550.12 nm much more severely than that of 1310 nm, thus resulting in its rapid key rate decline. When the launch power reaches -0.76 dBm, the key rates of both wavelengths are the same, and afterwards 1310 nm quantum signals display more advantage from the low level of Raman noise at 1310 nm. In addition, as we can see, when the power is greater than 2.0 dBm, 1550.12 nm could not generate any secure keys while 1310 nm could still performs well up to a power level of about 10 dBm, which corresponds to the Tbps level of classical communication. Similar results could be obtained for the QKD counter-propagation scenario. Consequently, we choose 1310 nm as the wavelength of quantum signals in our experiment, which not only allows one to achieve higher degrees of isolation in suppressing the linear crosstalk through low-cost coarse wavelength division multiplexer (CWDM), but also avoid nonlinear four-wave mixing (FWM) effects when multiple C-band classical channels are used, which may produce additional noise to a 1550.12-nm quantum channel \cite{Peters09,Eraerds10}.

\section{Co-propagation of QKD and four 64-QAM classical channels}
\begin{figure*}[tbh]
\centering
\includegraphics[width=0.8\linewidth]{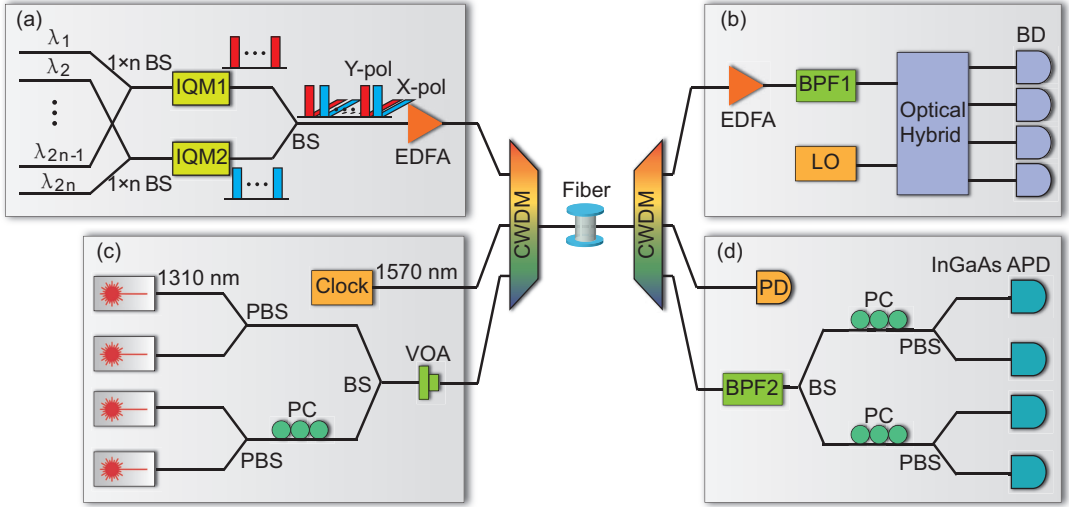}
\caption{Multiplexing schematic of QKD and Tbps data channels. (a) Classical transmitter. To simulate a real communication environment, we build up two sets of transmitters. The odd channels are combined by beam splitter (BS) to enter modulator 1, and the even channels are combined to enter modulator 2. Then the odd and even channels are combined in an interleaving way. The two in-phase and quadrature modulators (IQM) are driven by electrical signals with different data sequences, ensuring the adjacent channels carrying independent data. After an emulator of polarization division multiplexing, we use an EDFA to amplify and control the power of the classical light which enters the fibre link. (b) Classical receiver. At the receiver site, an EDFA amplifies the classical signals first, then a tunable bandpass filter at C-band (BPF1) is used to select the channel to be detected. In a polarization and phase diversity coherent receiver, the signal light and a local oscillator (LO) laser with approximately the same frequency are passed onto a polarization splitter and mixed in an optical hybrid, from which we operate balanced detection (BD) using paired photodiodes to accurately extract the signal amplitude and phase information. (c) Quantum transmitter. Four non-orthogonal states are generated through two polarizing beam splitters (PBS) and a polarization controller (PC), and the incident power of quantum states is adjusted by a variable optical attenuator (VOA). (d) Quantum receiver. We use a 100-GHz bandpass filter (BPF2) at 1310 nm to effectively suppress the Raman noise. The quantum signals are detected in two conjugate bases using InGaAs avalanche photodiode (APD) based single-photon detectors, and the states of polarization are controlled with automatic feedback.}
\label{fig:setup}
\end{figure*}

Figure~\ref{fig:setup} shows the experimental setup. Classical communication includes multiple DWDM channels within the C-band with wavelengths $\lambda_1$, $\lambda_2$, $\dots$, $\lambda_{2n-1}$, $\lambda_{2n}$. Meanwhile, our QKD system employs polarization encoding based BB84 protocol \cite{Bennett84} and the decoy state method against photon-number-splitting (PNS) attacks \cite{Hwang03,Lo05,Wang05,Ma05}. The clock synchronization between the QKD transmitter and receiver (referred to as Alice and Bob) is achieved with 100 kHz optical pulses at wavelength of 1570 nm. The classical, the quantum, and the synchronization channels are multiplexed and de-multiplexed using CWDMs to transmit over a single standard single-mode fibre. The CWDMs provide about 83-dB suppression of the in-band noise in the multiplexing and $>$180 dB isolation between the classical and the quantum channels in the de-multiplexing, which is sufficient to reduce the linear crosstalk to a negligible level. Before the detection of quantum signals, we use a custom-made 1310-nm bandpass filter with a bandwidth of 100 GHz, to diminish the Raman noise down to about 1/24 of that passes through the de-multiplexing CWDMs. The single-photon detectors can also effectively reduce the Raman noise in time domain through narrow gate widths.

In the first set of experiments, the classical optical communication system consists of 4 channels modulated with 64-QAM format. The channel spacing is 50 GHz with wavelengths ranging from 1549.1 to 1550.3 nm (the optical spectrum is shown as Supplementary Fig. 1). The bit rate of each channel is 336 Gbps, and thus the total gross data capacity is 1.344 Tbps. The co- and counter-propagating WDM layouts each induce a total loss of about 1.6 dB to classical channels (see Supplementary Fig. 2 and Fig. 3). Figure~\ref{fig:optimalPower} shows the measured classical bit error rate (BER) and Raman noise as functions of classical launch power after 50 km standard single-mode fibre (SSMF) transmission, with QKD co-propagating with classical channels in a WDM way. One can see that the classical BER is slightly higher with QKD than that of without, due to the additional attenuation induced by QKD multiplexing. The BER has a minimum value at 4-dBm launch power. As the power increases, the nonlinear distortions will degrade the signal quality and thus increase the BER. In addition, as shown in Fig.~\ref{fig:optimalPower}, the amount of Raman noise generated from the classical signal at 1310 nm is proportional to the incident light power, indicating that the spontaneous Raman scattering is a linear effect.

\begin{figure}[tbh]
\centering
\includegraphics[width=\linewidth]{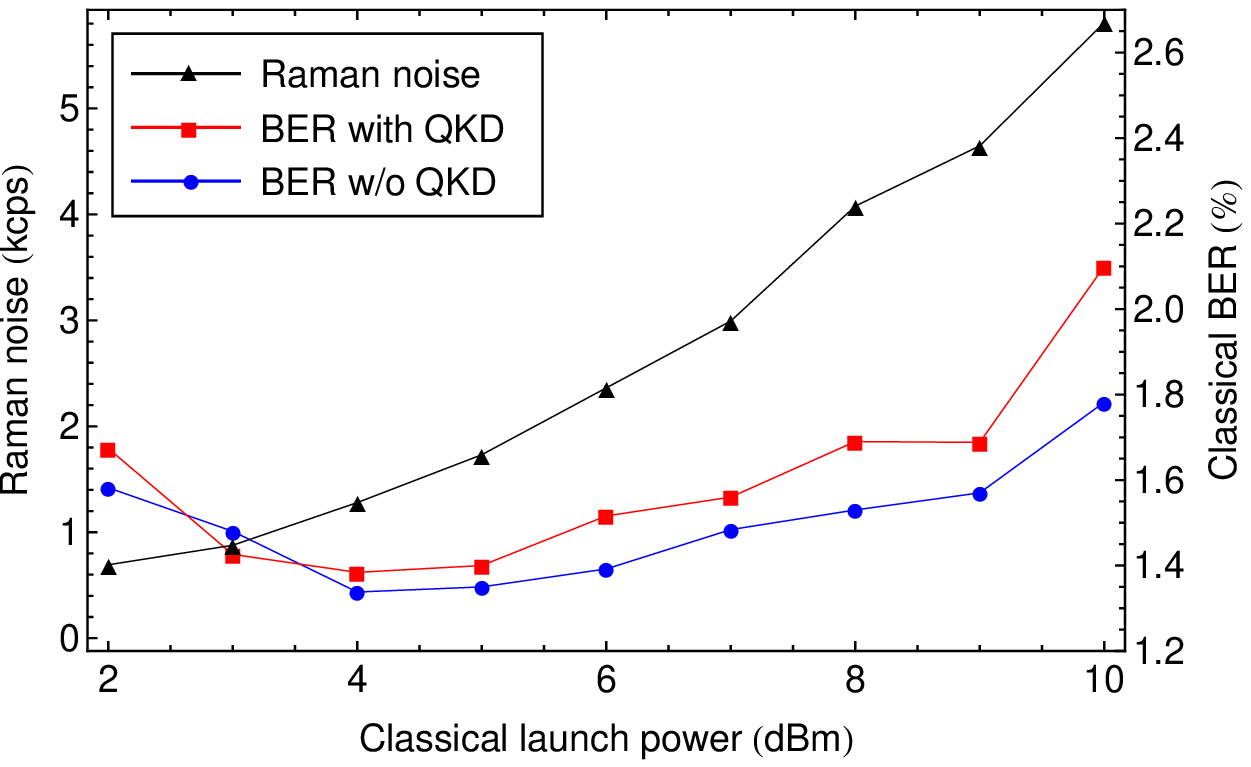}
\caption{The classical bit error rate (BER) and forward Raman noise (measured at 1310 nm) as functions of classical launch power at 50 km.}
\label{fig:optimalPower}
\end{figure}

\begin{figure}[tbh]
\centering
\includegraphics[width=\linewidth]{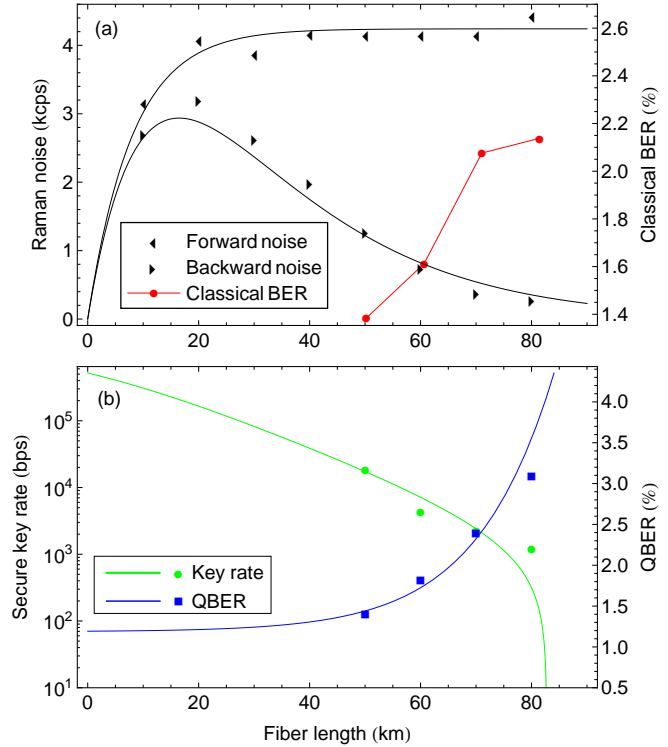}
\caption{Classical BER and QKD performance with WDM. (a) Measured (symbols) and simulated (solid lines) forward and backward Raman noise as a function of fibre length, and the measured classical BER (red dots) with WDM. (b) Measured and simulated QKD secure key rate (green dots and line) and QBER (blue squares and line), with quantum signals co-propagating with 4 classical channels.}
\label{fig:KeyRatevsLength}
\end{figure}

Figure~\ref{fig:KeyRatevsLength}(a) shows the Raman noise and the classical BER measured at different fibre distances in the WDM environment. Both the forward and the backward Raman noises are measured at 4-dBm launch power. As the transmission distance increases, the forward Raman noise first increases and then decreases, while the backward Raman noise increases gradually until saturation. Also, the backward noise count is much higher than the forward noise, which is consistent with theoretical calculations. It should be noted that, for classical communication forward error correction (FEC) is usually adopted, which can correct a pre-FEC BER of 0.45\% or 2.4\% to a level of $10^{-15}$ or less by adding hard- or soft-decision FEC with 7\% or 20\% overhead, respectively \cite{chang2010forward,chang2011fpga}.

Since Tbps communication is generally deployed in optical trunk links, we demonstrate the co-propagation of QKD and the four classical data channels at moderately longer distances. Figure~\ref{fig:KeyRatevsLength}(b) shows the QKD secure key rate and quantum bit error rate (QBER) in this scenario. The QKD secure key rate after 50 km transmission is 18.7 kbps, and the classical launch power is kept at 4 dBm from 50 km to 70 km with the BER below 2.4\% (see Supplementary Fig. 4). The maximum distance we achieved is 80 km with a fibre loss of 27.1 dB at 1310 nm, and the secure key rate is 1.2 kbps and QBER is 3.1\%. For 80 km co-propagation, we have to increase the optical power of the classical channels to 8 dBm in order to ensure its BER to be below 2.4\% (2.14\% in the experimental measurement, see Supplementary Fig. 5). Considering the soft-decision FEC with 20\% redundancy and frame overhead, the net bit rate of the classical communication is actually 1.07 Tbps. In the counter-propagating case, QKD suffers from much stronger backward Raman scattering. From Fig.~\ref{fig:KeyRatevsLength}(a) one can see that the backward Raman noise count at 50 km is 3.2 times more than its forward noise, resulting in QBER of 1.98\%, and key rate of 17.7 kbps. The maximum distance we achieved in the counter-propagating case is 70 km with QBER at 2.62\% and key rate of 3.7 kbps, where we have increased the classical launch power to 5 dBm with a measured BER of 2.18\% ( $<$2.4\%), and the net bit rate of the classical channels is still 1.07 Tbps.

\section{Co-propagation of QKD and 32 16-QAM classical channels}
\begin{figure}[tbh]
\centering
\includegraphics[width=\linewidth]{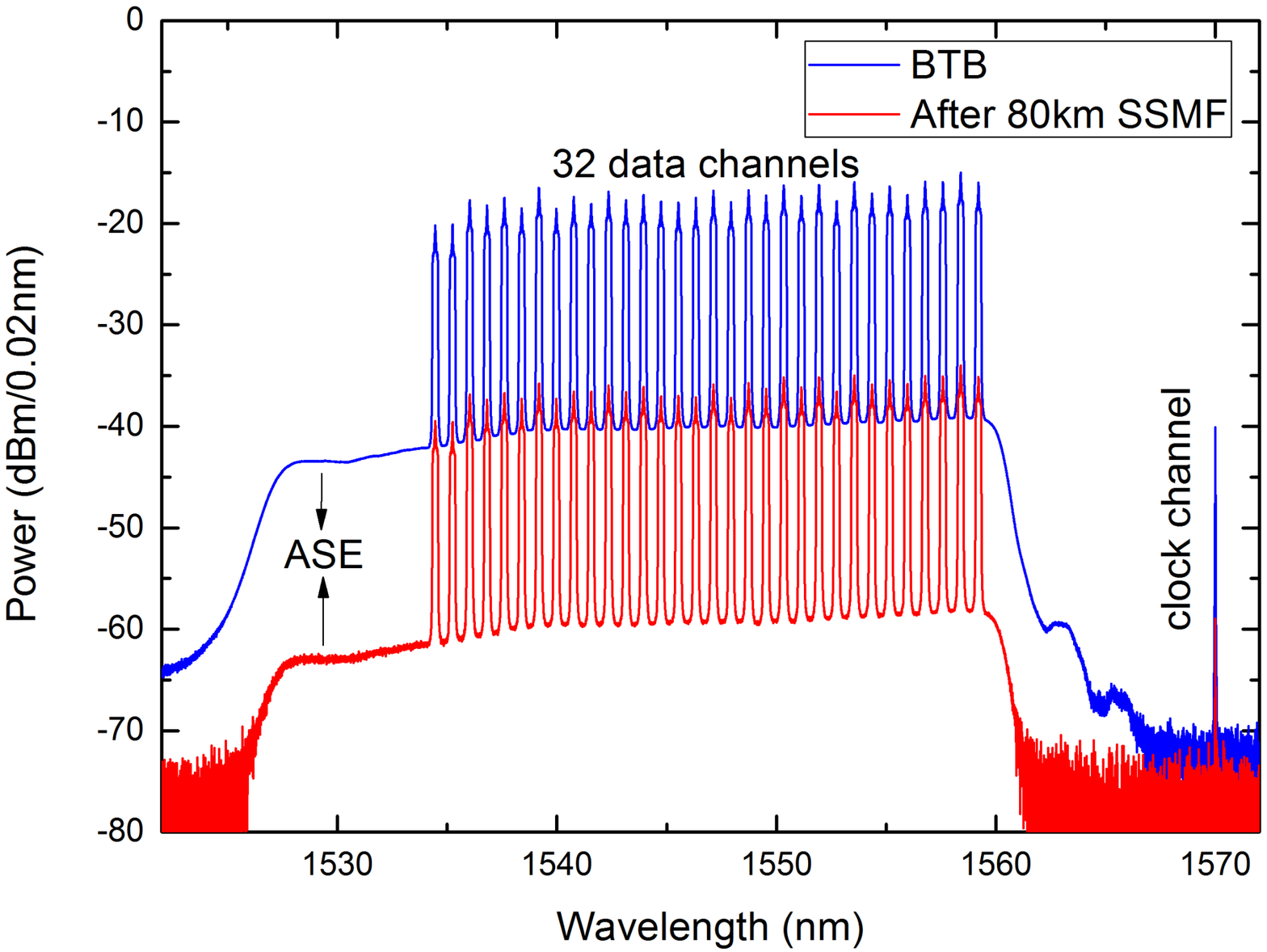}
\caption{The spectrum of 32 classical data channels and QKD clock synchronous channel, which is measured back-to-back (BTB) and after transmission over 80 km standard single mode fibre. One can see the obvious amplified spontaneous emission (ASE) generated from the erbium-doped fibre amplifier (EDFA) which should be suppress in advance to reduce the crosstalk.}
\label{fig:spectrum32channels}
\end{figure}

In the second set of experiments, we build up a classical optical communication system consisting of 32 channels modulated with a 16-QAM format. The channel spacing is 100 GHz with wavelengths ranging from 1535.7 to 1559.7 nm, and the optical spectrum is shown as Fig.~\ref{fig:spectrum32channels}. The bit rate of each channel is 224 Gbps, thus the total gross bandwidth amounts to 7.168 Tbps. The WDM layouts introduce about 2-dB loss to the classical channels (see Supplementary Fig. 6 and Fig. 7). We successfully implement the WDM of QKD and classical communication at different fibre distances for both co- and counter-propagating cases. In Table~\ref{tab:32channels} we list the measured results of 50 km and the maximum distance achievable.

\begin{table*}[]
\centering
\caption{The results of multiplexing QKD and 32 data channels.}
\label{tab:32channels}
\begin{ruledtabular}
\begin{tabular}{clcccd}
Direction  & \multicolumn{1}{c}{Distance (km)}  & BER    & Throughput (Tbps) & QBER   & \multicolumn{1}{c}{\textrm{Key Rate (kbps)}} \\
\hline
\multirow{2}{50pt}{Co-propagating}           & \quad 50       & 0.14\% & 6.38  & 1.48\% & 14.8  \\
                                             & \quad 80 (max) & 0.77\% & 5.69  & 4.24\% & 1.0   \\
\hline
\multirow{2}{50pt}{Counter-propagating}      & \quad 50       & 0.14\% & 6.38  & 2.08\% & 8.7   \\
                                             & \quad 60 (max) & 0.15\% & 6.38  & 3.24\% & 4.3   \\
\end{tabular}
\end{ruledtabular}
\end{table*}

We have measured that the optimal launch power for 50 km transmission is around 11 dBm (see Supplementary Fig. 8). We obtain the classical BER to be below 0.45\% when fibre distance is less than or equal to 70 km (see Supplementary Fig. 9), so we can perform error correction by adding 7\% overhead, therefore the effective throughput of classical channels reaches 6.38 Tbps, improving two orders of magnitude compared with previous results \cite{Patel14,choi2014field}. We achieve maximum transmission distances of 80 km and 60 km in the co- and counter-propagating cases, respectively.

\section{Conclution}
In our experiments, the WDM optical arrangements follow the principle of guaranteeing sufficient isolation of linear crosstalk, while using as few filters as possible, so as to reduce optical loss and cost. WDM filters generally have three ports: a common-port, a pass-port, and a reflect-port. We find that the pass-ports have much higher isolation than the reflect-ports. For instance, the pass-ports of 1550-nm (1310-nm) CWDMs have about 83-dB (90-dB) isolation to the light with wavelength 1310 nm (1550 nm), while the reflect-ports have an isolation of only about 20 dB for the filter center wavelength. Therefore, we use one 1550-nm CWDM to suppress the in-band noise, and two cascaded 1310-nm CWDMs to suppress the out-band noise. In addition, the second set of experiments have similar arrangements except that the 1550-nm CWDMs are replaced by 1550-nm filter-based wavelength division multiplexers (FWDMs), which have wider passband to accommodate all 32 channels.

For wavelength division multiplexing the main challenges of suppressing linear crosstalk and reducing Raman noise are irrelevant to the implemented QKD protocol and encoding format. Therefore, although we adopt BB84 protocol with polarization encoding in our experiment, the wavelength division multiplexing principle and methods we propose are adaptive to other QKD protocols, like differential phase shift QKD and measurement device independent QKD, and other encoding formats, like phase and time-bin encoding. Furthermore, our methods may also be used by continuous variable QKD or other kinds of quantum communications when they co-propagate with classical data channels over optical fibre.

The secure key rate and transmission distance are two important parameters of QKD, and are related to the performance of single-photon detectors and parameter estimation process. In our experiment, we use semi-conductor APD based detectors, but currently the superconducting nanowire single-photon detectors (SNSPDs) have better performances with detection efficiency of $>$70\% and dark count rate of $<$100 counts per second. Therefore, if the detector of our QKD system upgrades to SNSPD, the secure key rates and transmission distances of QKD will improve drastically. In addition, the finite key length we use to estimate parameters is $1 \times 10^6$. By increasing the statistical length we can obtain tighter parameter estimation, which would result in higher secure key rate and longer transmission distance.

In conclusion, we analyze the suitable wavelength for QKD transmission when multiplexed with C-band classical optical communication, and find that compared with 1550.12 nm, 1310 nm quantum signals are more adaptable in a Tbps classical data transmission environment with about 10-dBm launch power. Under this wavelength allocation, we have achieved more sufficient crosstalk isolation against classical channels using low-cost CWDMs. In addition, we have reduced the Raman noise through 100-GHz passband filters and single photon detectors with 180 ps gate width. Consequently, we demonstrate the wavelength division multiplexing of QKD with 16-QAM/64-QAM coherent optical communication, with a maximum throughput of 6.38 Tbps and a maximum transmission distance of 80 km, which is the typical span distance in classical communications. We note that although the secure key rate at 80 km is relatively low, the key rate at 50 km is still enough for voice and text encryption using one-time pad, and through using SNSPDs or trusted relays we can realize farther key distributions. Due to the high capacity of coherent optical communication, it will be a mainstream in future and may be applied in metropolitan and access networks, and thus QKD can be deployed in more classical optical communication environments and provide high security applications at low costs.

\begin{acknowledgments}
This work has been supported by the Science and Technological Fund of Anhui Province for Outstanding Youth (No. 1508085J02), the National Natural Science Foundation of China (No. 61475004) and the Chinese Academy of Sciences (No. XDA04030213). We thank Dan Wang for discussions.
\end{acknowledgments}

\appendix

\section{Classical communication subsystem}
In our experiments, the classical communication subsystem conveys multichannel WDM optical signals with digital Nyquist pulse shaping. The high order modulation formats such as 16-/64-QAM are adopted. We build two sets of transmitters and carry out transmission experiments of Terabit Nyquist polarization division multiplexing (PDM) 16/64-QAM signals (see Supplementary Methods). In our experiments, the arbitrary waveform generators (Keysight M8195A) operating at 56 GSample/s with 2-point DAC up-sampling generate baseband signals of 28 Gbaud. The digital RRC filters with a roll-off factor of 0.1 are chosen for Nyquist pulse shaping. We digitized and recorded the received data with a real-time oscilloscope (Keysight DSA-X 96204Q) for offline digital signal processing and signal quality evaluation. In the data frame of PDM Nyquist pulse shaping signal, the preamble of each polarization consist of synchronization and training sequences with a total length of 4696 symbols, which is followed by 102400 data symbols. Two pilot symbols are inserted in every 512 data symbols for carrier phase recovery. The data frame and the DSP diagrams of the transmitter and the receiver are detailed in Supplementary Fig. 11 and Fig. 12.

The bit error rate is used to signal quality evaluation. For each measurement, we record a total of $\sim10^6$ data symbols, that is, we evaluate $\sim4 \times 10^6$ received bits for 16-QAM signal and $\sim6 \times 10^6$ received bits for 64-QAM signal. To determine the BER, we perform the error counting by comparing the decoded symbols with the known bit sequence.

In our experiment, two raw BER criterions of $4.5 \times 10^{-3}$ and $2.4 \times 10^{-2}$ are adopted, which are the respective thresholds for error-free transmission when second-generation hard-decision FEC with 7\% overhead \cite{chang2010forward} or soft-decision FEC with 20\% overhead are used \cite{chang2011fpga}. If the FEC works properly, the corrected output BER is less than $1 \times 10^{-15}$, which can be considered as error-free transmission.

\section{Quantum key distribution subsystem}
Our QKD system operates at 625 MHz using polarization encoding. Alice encode the information using weak coherent laser sources and Bob detects the signals with InGaAs APD based single-photon detectors. The detectors work at gated mode with detection efficiency of 10\% and dark count rate of $1 \times 10^{-6}$ per clock cycle. To reduce the probability of afterpulsing, we set the dead time of detectors to 200 ns. In addition, we implement the decoy-state method to protect the system from potential photon-number-splitting attacks. Alice launches the signal states, decoy states, and vacuum states with a probability ratio of 6:1:1, and the average photon numbers of signal and decoy states are 0.6 and 0.2, respectively. We note that the vacuum state is also a kind of decoy state.

In the simulation, we estimate the single-photon parameters and calculate the QBER and secure key rate following the decoy state approach \cite{Lo05getting,Ma05}. The quantum bit error rate (QBER) is given by
\begin{eqnarray}
E_\mu=\frac{1}{Q_\mu}\left[\frac{1}{2}Y_0 +e_{opt}(1-Y_0)(1-e^{-\eta \mu})\right]
\end{eqnarray}
where $Q_{\mu}$ and $Y_0$ are the probabilities of an detection event given Alice emits a signal state and a vacuum state, respectively, $e_{opt}$ is the probability that a photon hit the erroneous detector due to finite polarization contrast, which is about 0.5\% for our system, and ¦Ç is the overall transmittance, including fibre loss, 3-dB loss of Bob's optical components, and the 10\% detection efficiency of single-photon detectors. In our experiment, the $Y_0$ contains two kinds of noise, including the dark count and afterpulsing of detectors, and the spontaneous Raman scattering from classical light. Meanwhile, the secure key rate per clock cycle is given by
\begin{eqnarray}
R= q\{ -Q_\mu f H_2(E_\mu) + Q_1[1-H_2(e_1)]+ Q_0\}
\end{eqnarray}
where $q$ is the probability of Alice emits signal states and Alice and Bob choose the same bases, $f$ is the inefficiency of error correction which is about 1.25, $e_1$ is the estimated error rate of single-photon states, and $Q_1$ and $Q_0$ are the fractions of detection events by Bob that is due to the single-photon and vacuum ingredients of signal states, respectively. $H_2(x)=-xlog_2(x)-(1-x)log_2(1-x)$ is the binary entropy function. The data block size we used to estimate parameters is $1 \times 10^6$, and we consider the statistical fluctuations of 5 standard derivations. In the experiment, we implement the entire QKD postprocessing \cite{fung2010practical} based on hardware, including message authentication with pre-shared symmetric keys, error correction with a cascade algorithm \cite{brassard1994secret}, error verification with a cyclic redundancy check (CRC), and privacy amplification with a Toeplitz matrix \cite{bennett1995generalized}.

\bibliographystyle{naturemag}
\bibliography{TeraWDMref}

\begin{thebibliography}{10}
\expandafter\ifx\csname url\endcsname\relax
  \def\url#1{\texttt{#1}}\fi
\expandafter\ifx\csname urlprefix\endcsname\relax\def\urlprefix{URL }\fi
\providecommand{\bibinfo}[2]{#2}
\providecommand{\eprint}[2][]{\url{#2}}

\bibitem{Bennett84}
\bibinfo{author}{Bennett, C.~H.} \& \bibinfo{author}{Brassard, G.}
\newblock {Quantum cryptography: public key distribution and coin tossing}.
\newblock In \emph{\bibinfo{booktitle}{Proceedings of the IEEE International
  Conference on Computers, Systems and Signal Processing, Bangalore, India}},
  \bibinfo{pages}{175--179} (\bibinfo{publisher}{IEEE}, \bibinfo{address}{New
  York}, \bibinfo{year}{1984}).

\bibitem{EKERT91}
\bibinfo{author}{Ekert, A.~K.}
\newblock {Quantum cryptography based on bell theorem}.
\newblock \emph{\bibinfo{journal}{{Phys. Rev. Lett.}}}
  \textbf{\bibinfo{volume}{{67}}}, \bibinfo{pages}{{661}}
  (\bibinfo{year}{{1991}}).

\bibitem{Gisin02}
\bibinfo{author}{Gisin, N.}, \bibinfo{author}{Ribordy, G.},
  \bibinfo{author}{Tittel, W.} \& \bibinfo{author}{Zbinden, H.}
\newblock {Quantum cryptography}.
\newblock \emph{\bibinfo{journal}{{Rev. Mod. Phys.}}}
  \textbf{\bibinfo{volume}{{74}}}, \bibinfo{pages}{{145}}
  (\bibinfo{year}{{2002}}).

\bibitem{Scarani09}
\bibinfo{author}{Scarani, V.} \emph{et~al.}
\newblock {The security of practical quantum key distribution}.
\newblock \emph{\bibinfo{journal}{{Rev. Mod. Phys.}}}
  \textbf{\bibinfo{volume}{{81}}}, \bibinfo{pages}{{1301}}
  (\bibinfo{year}{{2009}}).

\bibitem{Takesue07}
\bibinfo{author}{Takesue, H.} \emph{et~al.}
\newblock {Quantum key distribution over a 40-dB channel loss using
  superconducting single-photon detectors}.
\newblock \emph{\bibinfo{journal}{{Nat. Photon.}}}
  \textbf{\bibinfo{volume}{{1}}}, \bibinfo{pages}{{343}}
  (\bibinfo{year}{{2007}}).

\bibitem{Stucki09}
\bibinfo{author}{Stucki, D.} \emph{et~al.}
\newblock {High rate, long-distance quantum key distribution over 250 km of
  ultra low loss fibres}.
\newblock \emph{\bibinfo{journal}{{New J. Phys.}}}
  \textbf{\bibinfo{volume}{{11}}}, \bibinfo{pages}{{075003}}
  (\bibinfo{year}{{2009}}).

\bibitem{Dixon10}
\bibinfo{author}{Dixon, A.~R.}, \bibinfo{author}{Yuan, Z.~L.},
  \bibinfo{author}{Dynes, J.~F.}, \bibinfo{author}{Sharpe, A.~W.} \&
  \bibinfo{author}{Shields, A.~J.}
\newblock {Continuous operation of high bit rate quantum key distribution}.
\newblock \emph{\bibinfo{journal}{{Appl. Phys. Lett.}}}
  \textbf{\bibinfo{volume}{{96}}}, \bibinfo{pages}{{161102}}
  (\bibinfo{year}{{2010}}).

\bibitem{Liu10}
\bibinfo{author}{Liu, Y.} \emph{et~al.}
\newblock {Decoy-state quantum key distribution with polarized photons over 200
  km}.
\newblock \emph{\bibinfo{journal}{{Opt. Express}}}
  \textbf{\bibinfo{volume}{{18}}}, \bibinfo{pages}{{8587}}
  (\bibinfo{year}{{2010}}).

\bibitem{Wang12}
\bibinfo{author}{Wang, S.} \emph{et~al.}
\newblock {2 GHz clock quantum key distribution over 260 km of standard telecom
  fiber}.
\newblock \emph{\bibinfo{journal}{{Opt. Lett.}}}
  \textbf{\bibinfo{volume}{{37}}}, \bibinfo{pages}{{1008}}
  (\bibinfo{year}{{2012}}).

\bibitem{Tanaka12}
\bibinfo{author}{Tanaka, A.} \emph{et~al.}
\newblock {High-speed quantum key distribution system for 1-Mbps real-time key
  generation}.
\newblock \emph{\bibinfo{journal}{{IEEE J. Quantum Electron.}}}
  \textbf{\bibinfo{volume}{{48}}}, \bibinfo{pages}{{542}}
  (\bibinfo{year}{{2012}}).

\bibitem{korzh2015provably}
\bibinfo{author}{Korzh, B.} \emph{et~al.}
\newblock Provably secure and practical quantum key distribution over 307 km of
  optical fibre.
\newblock \emph{\bibinfo{journal}{Nat. Photonics}}
  \textbf{\bibinfo{volume}{9}}, \bibinfo{pages}{163--168}
  (\bibinfo{year}{2015}).

\bibitem{Peev09}
\bibinfo{author}{Peev, M.} \emph{et~al.}
\newblock {The SECOQC quantum key distribution network in Vienna}.
\newblock \emph{\bibinfo{journal}{{New J. Phys.}}}
  \textbf{\bibinfo{volume}{{11}}}, \bibinfo{pages}{{075001}}
  (\bibinfo{year}{{2009}}).

\bibitem{Chen10}
\bibinfo{author}{Chen, T.~Y.} \emph{et~al.}
\newblock {Metropolitan all-pass and inter-city quantum communication network}.
\newblock \emph{\bibinfo{journal}{{Opt. Express}}}
  \textbf{\bibinfo{volume}{{18}}}, \bibinfo{pages}{{27217}}
  (\bibinfo{year}{{2010}}).

\bibitem{Sasaki11}
\bibinfo{author}{Sasaki, M.} \emph{et~al.}
\newblock {Field test of quantum key distribution in the Tokyo QKD Network}.
\newblock \emph{\bibinfo{journal}{{Opt. Express}}}
  \textbf{\bibinfo{volume}{{19}}}, \bibinfo{pages}{{10387}}
  (\bibinfo{year}{{2011}}).

\bibitem{Froehlich13}
\bibinfo{author}{Fr\"{o}hlich, B.} \emph{et~al.}
\newblock {A quantum access network}.
\newblock \emph{\bibinfo{journal}{{Nature}}} \textbf{\bibinfo{volume}{{501}}},
  \bibinfo{pages}{{69--72}} (\bibinfo{year}{{2013}}).

\bibitem{Townsend97}
\bibinfo{author}{Townsend, P.~D.}
\newblock {Simultaneous quantum cryptographic key distribution and conventional
  data transmission over installed fibre using wavelength-division
  multiplexing}.
\newblock \emph{\bibinfo{journal}{{Electron. Lett.}}}
  \textbf{\bibinfo{volume}{{33}}}, \bibinfo{pages}{{188}}
  (\bibinfo{year}{{1997}}).

\bibitem{Chapuran09}
\bibinfo{author}{Chapuran, T.~E.} \emph{et~al.}
\newblock {Optical networking for quantum key distribution and quantum
  communications}.
\newblock \emph{\bibinfo{journal}{{New J. Phys.}}}
  \textbf{\bibinfo{volume}{{11}}}, \bibinfo{pages}{{105001}}
  (\bibinfo{year}{{2009}}).

\bibitem{Choi11}
\bibinfo{author}{Choi, I.}, \bibinfo{author}{Young, R.~J.} \&
  \bibinfo{author}{Townsend, P.~D.}
\newblock {Quantum information to the home}.
\newblock \emph{\bibinfo{journal}{{New J. Phys.}}}
  \textbf{\bibinfo{volume}{{13}}}, \bibinfo{pages}{{063039}}
  (\bibinfo{year}{{2011}}).

\bibitem{Choi10}
\bibinfo{author}{Choi, I.}, \bibinfo{author}{Young, R.~J.} \&
  \bibinfo{author}{Townsend, P.~D.}
\newblock {Quantum key distribution on a 10Gb/s WDM-PON}.
\newblock \emph{\bibinfo{journal}{{Opt. Express}}}
  \textbf{\bibinfo{volume}{{18}}}, \bibinfo{pages}{{9600}}
  (\bibinfo{year}{{2010}}).

\bibitem{Peters09}
\bibinfo{author}{Peters, N.~A.} \emph{et~al.}
\newblock {Dense wavelength multiplexing of 1550nm QKD with strong classical
  channels in reconfigurable networking environments}.
\newblock \emph{\bibinfo{journal}{{New J. Phys.}}}
  \textbf{\bibinfo{volume}{{11}}}, \bibinfo{pages}{045012}
  (\bibinfo{year}{{2009}}).

\bibitem{lancho2009qkd}
\bibinfo{author}{Lancho, D.}, \bibinfo{author}{Martinez, J.},
  \bibinfo{author}{Elkouss, D.}, \bibinfo{author}{Soto, M.} \&
  \bibinfo{author}{Martin, V.}
\newblock QKD in standard optical telecommunications networks.
\newblock In \emph{\bibinfo{booktitle}{Quantum Communication and Quantum
  Networking}}, \bibinfo{pages}{142--149} (\bibinfo{publisher}{Springer},
  \bibinfo{year}{2009}).

\bibitem{mora2012simultaneous}
\bibinfo{author}{Mora, J.} \emph{et~al.}
\newblock Simultaneous transmission of 20x2 WDM/SCM-QKD and 4 bidirectional
  classical channels over a PON.
\newblock \emph{\bibinfo{journal}{Opt. Express}} \textbf{\bibinfo{volume}{20}},
  \bibinfo{pages}{16358--16365} (\bibinfo{year}{2012}).

\bibitem{Aleksic13}
\bibinfo{author}{Aleksic, S.} \emph{et~al.}
\newblock {Quantum key distribution over optical access networks}.
\newblock In \emph{\bibinfo{booktitle}{{Proceedings of the 18th European
  Conference on Network and Optical Communications, Graz, Austria}}},
  \bibinfo{pages}{{11--18}} (\bibinfo{publisher}{IEEE}, \bibinfo{address}{New
  York}, \bibinfo{year}{{2013}}).

\bibitem{Ciurana14}
\bibinfo{author}{Ciurana, A.} \emph{et~al.}
\newblock {Quantum metropolitan optical network based on wavelength division
  multiplexing}.
\newblock \emph{\bibinfo{journal}{{Opt. Express}}}
  \textbf{\bibinfo{volume}{{22}}}, \bibinfo{pages}{{1576}}
  (\bibinfo{year}{{2014}}).

\bibitem{aleksic2015perspectives}
\bibinfo{author}{Aleksic, S.} \emph{et~al.}
\newblock Perspectives and limitations of QKD integration in metropolitan area
  networks.
\newblock \emph{\bibinfo{journal}{Opt. Express}} \textbf{\bibinfo{volume}{23}},
  \bibinfo{pages}{10359--10373} (\bibinfo{year}{2015}).

\bibitem{Patel14}
\bibinfo{author}{Patel, K.~A.} \emph{et~al.}
\newblock {Quantum key distribution for 10 Gb/s dense wavelength division
  multiplexing networks}.
\newblock \emph{\bibinfo{journal}{{Appl. Phys. Lett.}}}
  \textbf{\bibinfo{volume}{{104}}}, \bibinfo{pages}{051123}
  (\bibinfo{year}{{2014}}).

\bibitem{choi2014field}
\bibinfo{author}{Choi, I.} \emph{et~al.}
\newblock Field trial of a quantum secured 10 Gb/s DWDM transmission system
  over a single installed fiber.
\newblock \emph{\bibinfo{journal}{Opt. Express}} \textbf{\bibinfo{volume}{22}},
  \bibinfo{pages}{23121--23128} (\bibinfo{year}{2014}).

\bibitem{auyeung1978spontaneous}
\bibinfo{author}{Auyeung, J.} \& \bibinfo{author}{Yariv, A.}
\newblock Spontaneous and stimulated Raman scattering in long low loss fibers.
\newblock \emph{\bibinfo{journal}{IEEE J. Quant. Electron.}}
  \textbf{\bibinfo{volume}{14}}, \bibinfo{pages}{347--352}
  (\bibinfo{year}{1978}).

\bibitem{Subacius05}
\bibinfo{author}{Subacius, D.}, \bibinfo{author}{Zavriyev, A.} \&
  \bibinfo{author}{Trifonov, A.}
\newblock {Backscattering limitation for fiber-optic quantum key distribution
  systems}.
\newblock \emph{\bibinfo{journal}{{Appl. Phys. Lett.}}}
  \textbf{\bibinfo{volume}{{86}}}, \bibinfo{pages}{011103}
  (\bibinfo{year}{{2005}}).

\bibitem{xavier2009scattering}
\bibinfo{author}{Xavier, G.~B.}, \bibinfo{author}{de~Faria, G.~V.},
  \bibinfo{author}{Tempor{\~a}o, G.~P.} \& \bibinfo{author}{von~der Weid,
  J.~P.}
\newblock {Scattering effects on QKD employing simultaneous classical and
  quantum channels in telecom optical fibers in the C-band}.
\newblock In \emph{\bibinfo{booktitle}{Quantum Communication, Measurement and
  Computing (QCMC): Ninth International Conference on QCMC}},
  \bibinfo{pages}{327--330} (\bibinfo{organization}{AIP Publishing},
  \bibinfo{year}{2009}).

\bibitem{da2014impact}
\bibinfo{author}{da~Silva, T.~F.}, \bibinfo{author}{Xavier, G.~B.},
  \bibinfo{author}{Tempor{\~a}o, G.~P.} \& \bibinfo{author}{von~der Weid,
  J.~P.}
\newblock Impact of Raman scattered noise from multiple telecom channels on
  fiber-optic quantum key distribution systems.
\newblock \emph{\bibinfo{journal}{J. Lightwave Technol.}}
  \textbf{\bibinfo{volume}{32}}, \bibinfo{pages}{2332--2339}
  (\bibinfo{year}{2014}).

\bibitem{wootters1982single}
\bibinfo{author}{Wootters, W.~K.} \& \bibinfo{author}{Zurek, W.~H.}
\newblock A single quantum cannot be cloned.
\newblock \emph{\bibinfo{journal}{Nature}} \textbf{\bibinfo{volume}{299}},
  \bibinfo{pages}{802--803} (\bibinfo{year}{1982}).

\bibitem{dieks1982communication}
\bibinfo{author}{Dieks, D.}
\newblock Communication by EPR devices.
\newblock \emph{\bibinfo{journal}{Physics Letters A}}
  \textbf{\bibinfo{volume}{92}}, \bibinfo{pages}{271--272}
  (\bibinfo{year}{1982}).

\bibitem{xia2006band}
\bibinfo{author}{Xia, T.~J.} \emph{et~al.}
\newblock In-band quantum key distribution (QKD) on fiber populated by
  high-speed classical data channels.
\newblock In \emph{\bibinfo{booktitle}{Optical Fiber Communication
  Conference}}, \bibinfo{pages}{OTuJ7} (\bibinfo{organization}{Optical Society
  of America}, \bibinfo{year}{2006}).

\bibitem{Eraerds10}
\bibinfo{author}{Eraerds, P.}, \bibinfo{author}{Walenta, N.},
  \bibinfo{author}{Legr\'{e}, M.}, \bibinfo{author}{Gisin, N.} \&
  \bibinfo{author}{Zbinden, H.}
\newblock {Quantum key distribution and 1 Gbps data encryption over a single
  fibre}.
\newblock \emph{\bibinfo{journal}{{New J. Phys.}}}
  \textbf{\bibinfo{volume}{{12}}}, \bibinfo{pages}{063027}
  (\bibinfo{year}{{2010}}).

\bibitem{Patel12}
\bibinfo{author}{Patel, K.~A.} \emph{et~al.}
\newblock {Coexistence of high-bit-rate quantum key distribution and data on
  optical fiber}.
\newblock \emph{\bibinfo{journal}{{Phys. Rev. X}}}
  \textbf{\bibinfo{volume}{{2}}}, \bibinfo{pages}{041010}
  (\bibinfo{year}{{2012}}).

\bibitem{walenta2014fast}
\bibinfo{author}{Walenta, N.} \emph{et~al.}
\newblock A fast and versatile quantum key distribution system with hardware
  key distillation and wavelength multiplexing.
\newblock \emph{\bibinfo{journal}{New J. Phys.}} \textbf{\bibinfo{volume}{16}},
  \bibinfo{pages}{013047} (\bibinfo{year}{2014}).

\bibitem{wang2015experimental}
\bibinfo{author}{Wang, L.-J.} \emph{et~al.}
\newblock Experimental multiplexing of quantum key distribution with classical
  optical communication.
\newblock \emph{\bibinfo{journal}{Appl. Phys. Lett.}}
  \textbf{\bibinfo{volume}{106}}, \bibinfo{pages}{081108}
  (\bibinfo{year}{2015}).

\bibitem{runser2005demonstration}
\bibinfo{author}{Runser, R.~J.} \emph{et~al.}
\newblock Demonstration of 1.3 $\mu$m quantum key distribution (QKD)
  compatibility with 1.5 $\mu$m metropolitan wavelength division multiplexed
  (WDM) systems.
\newblock In \emph{\bibinfo{booktitle}{Optical Fiber Communication
  Conference}}, \bibinfo{pages}{OWI2} (\bibinfo{organization}{Optical Society
  of America}, \bibinfo{year}{2005}).

\bibitem{Nweke05}
\bibinfo{author}{Nweke, N.~I.} \emph{et~al.}
\newblock {Experimental characterization of the separation between
  wavelength-multiplexed quantum and classical communication channels}.
\newblock \emph{\bibinfo{journal}{{Appl. Phys. Lett.}}}
  \textbf{\bibinfo{volume}{{87}}}, \bibinfo{pages}{{174103}}
  (\bibinfo{year}{{2005}}).

\bibitem{Ma05}
\bibinfo{author}{Ma, X.~F.}, \bibinfo{author}{Qi, B.}, \bibinfo{author}{Zhao,
  Y.} \& \bibinfo{author}{Lo, H.~K.}
\newblock {Practical decoy state for quantum key distribution}.
\newblock \emph{\bibinfo{journal}{{Phys. Rev. A}}}
  \textbf{\bibinfo{volume}{{72}}}, \bibinfo{pages}{012326}
  (\bibinfo{year}{{2005}}).

\bibitem{Hwang03}
\bibinfo{author}{Hwang, W.~Y.}
\newblock {Quantum key distribution with high loss: toward global secure
  communication}.
\newblock \emph{\bibinfo{journal}{{Phys. Rev. Lett.}}}
  \textbf{\bibinfo{volume}{{91}}}, \bibinfo{pages}{057901}
  (\bibinfo{year}{{2003}}).

\bibitem{Lo05}
\bibinfo{author}{Lo, H.~K.}, \bibinfo{author}{Ma, X.~F.} \&
  \bibinfo{author}{Chen, K.}
\newblock {Decoy state quantum key distribution}.
\newblock \emph{\bibinfo{journal}{{Phys. Rev. Lett.}}}
  \textbf{\bibinfo{volume}{{94}}}, \bibinfo{pages}{230504}
  (\bibinfo{year}{{2005}}).

\bibitem{Wang05}
\bibinfo{author}{Wang, X.~B.}
\newblock {Beating the photon-number-splitting attack in practical quantum
  cryptography}.
\newblock \emph{\bibinfo{journal}{{Phys. Rev. Lett.}}}
  \textbf{\bibinfo{volume}{{94}}}, \bibinfo{pages}{230503}
  (\bibinfo{year}{{2005}}).

\bibitem{chang2010forward}
\bibinfo{author}{Chang, F.}, \bibinfo{author}{Onohara, K.} \&
  \bibinfo{author}{Mizuochi, T.}
\newblock Forward error correction for 100 G transport networks.
\newblock \emph{\bibinfo{journal}{Communications Magazine, IEEE}}
  \textbf{\bibinfo{volume}{48}}, \bibinfo{pages}{S48--S55}
  (\bibinfo{year}{2010}).

\bibitem{chang2011fpga}
\bibinfo{author}{Chang, D.} \emph{et~al.}
\newblock FPGA verification of a single QC-LDPC code for 100 Gb/s optical
  systems without error floor down to BER of 10\textsuperscript{-15}.
\newblock In \emph{\bibinfo{booktitle}{Optical Fiber Communication
  Conference}}, \bibinfo{pages}{OTuN2} (\bibinfo{organization}{Optical Society
  of America}, \bibinfo{year}{2011}).

\bibitem{Lo05getting}
\bibinfo{author}{Lo, H.-K.}
\newblock Getting something out of nothing.
\newblock \emph{\bibinfo{journal}{Quantum Info. Comput.}}
  \textbf{\bibinfo{volume}{5}}, \bibinfo{pages}{413--418}
  (\bibinfo{year}{2005}).

\bibitem{fung2010practical}
\bibinfo{author}{Fung, C.-H.~F.}, \bibinfo{author}{Ma, X.} \&
  \bibinfo{author}{Chau, H.}
\newblock Practical issues in quantum-key-distribution postprocessing.
\newblock \emph{\bibinfo{journal}{Phys. Rev. A}} \textbf{\bibinfo{volume}{81}},
  \bibinfo{pages}{012318} (\bibinfo{year}{2010}).

\bibitem{brassard1994secret}
\bibinfo{author}{Brassard, G.} \& \bibinfo{author}{Salvail, L.}
\newblock Secret-key reconciliation by public discussion.
\newblock In \bibinfo{editor}{Helleseth, T.} (ed.)
  \emph{\bibinfo{booktitle}{Advances in Cryptology --- EUROCRYPT '93: Workshop
  on the Theory and Application of Cryptographic Techniques Lofthus, Norway,
  May 23--27, 1993 Proceedings}}, \bibinfo{pages}{410--423}
  (\bibinfo{publisher}{Springer}, \bibinfo{address}{Berlin, Heidelberg},
  \bibinfo{year}{1994}).

\bibitem{bennett1995generalized}
\bibinfo{author}{Bennett, C.~H.}, \bibinfo{author}{Brassard, G.},
  \bibinfo{author}{Crepeau, C.} \& \bibinfo{author}{Maurer, U.~M.}
\newblock Generalized privacy amplification.
\newblock \emph{\bibinfo{journal}{IEEE Trans. Inf. Theory}}
  \textbf{\bibinfo{volume}{41}}, \bibinfo{pages}{1915--1923}
  (\bibinfo{year}{1995}).

\end{thebibliography}

\end{document}